\documentclass[12pt]{iopart}
\begin{document}

\title{High Energy Nucleus-Nucleus Collisions In View Of The Multi-Peripheral Model}
\author{M. T. Hussein and N. M. Hassan and $^{*}$ N. Elharby}
\address{ Physics Department, Faculty of Science, Cairo University, Cairo - Egypt}
\address{$^{*}$ Physics Department, Faculty of Education for Girls, Jeddah, KSA.}

\begin{abstract}
The hypothesis of the multi peripheral model is extended to the
hadron-nucleus interactions and then generalized to the nucleus-nucleus
case. The processing of the model depends on input parameters that are
extracted from the features of the experiments in this field. The number of
encountered nucleons from both target and projectile are estimated according
to the eikonal scattering approach. The screening effect due to the
interaction of the projectile nucleons in successive manner with the target
nucleus is considered. The rapidity distributions of fast particles are
reproduced for the successive collisions in p-S and $^{32}$S-$^{32}$S
interactions at 200 A GeV. A global fair agreement is found in comparison
with data of the experiment SLAC-NA-035.

\end{abstract}

\pacs{05.70.Cd}


\maketitle

\section{Introduction}

In the last few years the research work was concentrated on possible
existence of the quark-gluon plasma phase, considering of unconfined quarks
and gluons at high temperature or high density. In the laboratory,
nucleus-nucleus collisions at very high energies provide a promising way to
produce high temperature or high-density matter. As estimated by Bajorken
[1] the energy density can be so high that these reactions might be utilized
to explore the existence of the quark-gluon plasma. One of the many factors
that lead to an optimistic assessment that matter at high density and high
temperature may be produced with nucleus-nucleus collisions is the
occurrence of multiple collisions. By this mean, a nucleon of one nucleus
may collide with many nucleons in the other and deposits a large amount of
energy in the collision region. In the nucleon-nucleon center of mass
system, the longitudinal inter-nucleon spacing between target nucleons are
Lorentz contracted and can be smaller than 1 fm in high-energy collisions.
On the other hand, particle production occurs only when a minimum distance
of about 1 fm separates the leading quark and antiquark in the
nucleon-nucleon center of mass system [2,3]. Therefore when the projectile
nucleon collides with many target nucleons, particles production arising
from the first N-N collision is not finished before the collision of the
projectile with another target nucleon begins. There are models [4-11] that
describe how is the second collision is affected by the first one.
Nevertheless, the fundamental theory of doing that remains one of the
unsolved problems. Experimental data suggest that after the projectile
nucleon makes a collision, the projectile-like object that emerges from the
first collision appears to continue to collide with other nucleons in the
target nucleus on its way through the target nucleus. In each collision the
object that emerges along the projectile nucleon direction has a net baryon
number of unity because of the conservation of the baryon number. One can
speak loosely of this object as the projectile or baryon-like object and can
describe the multiple collision process in terms of the projectile nucleon
making many collisions with the target nucleons. Then, losing energy and
momentum in the process, and emerging from the other side of the target with
a much diminished energy. The number of collisions depends on the thickness
of the target nucleus. Experimental evidence of occurrence of multiple
collision process can be best illustrated with the data of p-A reactions in
the projectile fragmentation region [3, 12]. In the present work we shall
investigate the particle production mechanism in heavy ion collision by
extending the multi-peripheral model [13-15] to the nucleon-Nucleus and then
generalize to the nucleus-nucleus case. The paper is organized so that in
section {\bf 2}, we present the experimental features of high-energy events.
This will serve in setting the boundary condition of the extended problem
and building up the model in a realistic form. The hypothesis of the
multi-peripheral model is demonstrated in section{\bf \ 3}. The
nucleon-nucleon case is briefly described in {\bf 3.1} and the extension to
the nucleon-nucleus and the nucleus-nucleus collisions are followed in
section {\bf 3.2} and {\bf 3.3} respectively. Concluding remarks are given
in section {\bf 4}.

\section{Experimental Features of High Energy Events}

The heavy ion collisions initiate the chance of processing multiple
collisions of single nucleon-nucleon (N-N) events. This behavior allows the
magnification of any small signal that may appear unclear in (N-N) event.
The magnification increases as the size of the interacting nuclei increases.
Nevertheless, the relation is not simply linear but many other factors are
expected to contribute the complexity of the problem. The study of the
multiplicity distribution at different rapidity intervals carries valuable
information in this sense [16]. The rapidity range is classified into
regions that characterizes the type of the interaction. The target and the
projectile fragmentation regions contain the fast particles characterizing
the forward and backward production from the fragmentation of both the
target and the projectile nuclei in their center of mass system. The central
rapidity interval is the hot region of the reaction. It sends global
information about the strong interactions inside the nuclear bowl. Moreover,
signals about (QGP) may be extracted from the study of this region. Fig. (1)
shows the experimental data of the rapidity distributions of fast particles
produced in the interaction of sulfur-projectile with proton, sulfur, silver
and gold as target nuclei at 200 A GeV incident energy [17]. The average
multiplicity at each rapidity interval depends on the incident projectile
energy (which is fixed in this case) and the size of the interacting nuclei.
The nucleus-nucleus (A-A) collision may be seen as N-N base or N-A base. To
put the problem in a quantitative measurable form, we define a
multiplication factor {\it MF (N\_N)} as a base of (N-N);

\begin{equation}
MF_{(N-N)}=\frac{Yield, in(A-A)}{Yield, in (N-N)}
\end{equation}

And the multiplication factor MF (N-A) as a base of (N-A);

\begin{equation}
MF_{(N-A)}=\frac{Yield, in(A-A)}{Yield, in(N-A)}
\end{equation}

Fig. (2) and Fig. (3) show the multiplication factors based on (N-N) and
(N-A) for the different reactions. The multiplication factor has maximum
value at the central rapidity region (Y\symbol{126}0) in both cases. The
experimental data of p-S and the S-S collisions show constant value allover
the rapidity range. Nevertheless, it is difficult to estimate the number of
base-collisions in each reaction from this approach. An alternative approach
is to renormalize the rapidity distribution to get a scaled function that is
independent on the projectile and target size. The average number of
encountered nucleons from the target by an incident hadron [16,18] is a good
measure of the number of collisions inside a target nucleus. This is defined
as,

\begin{equation}
\nu =A_{t}\frac{\sigma _{hN}}{\sigma _{hA}}
\end{equation}

Where $\sigma _{hN}$, and $\sigma _{hA}$ are the total inelastic cross
section of (h-N) and (N-A) respectively. By analogy, the average number of
binary (N-N) collisions inside the (A-A) collision may be defined as,

\begin{equation}
\nu _{NN}=A_{p}A_{t}\frac{\sigma _{hN}}{\sigma _{AA}}
\end{equation}

$\sigma _{AA}$ is the total inelastic cross section of (A-A) collision at
the same incident energy. On the other hand, the average number of
collisions based to N-A is defined as,

\begin{equation}
\nu _{NA}=A_{p}\frac{\sigma _{hAt}}{\sigma _{AA}}
\end{equation}

is total inelastic cross section of (h-At) collision. The scaling function
is then obtained by dividing the rapidity distribution by the average number
of collisions or . The resultant are displayed in Fig. (4) which shows the
overlapped distributions corresponding to the reactions of different nuclear
sizes. Following now a geometric approach which assumes that the nuclear
radius is linearly proportional to the one third power of its mass number,
then Eqs.( 3-5) becomes,

\[
\nu =A_{t}^{1/3} 
\]

\begin{equation}
\nu _{NN}=\frac{A_{p}A_{t}}{A_{t}^{2/3}+A_{p}^{2/3}+2A_{t}^{1/3}A_{p}^{1/3}}
\end{equation}

\[
\nu _{NA}=\frac{A_{p}A_{t}^{2/3}}{%
A_{t}^{2/3}+A_{p}^{2/3}+2A_{t}^{1/3}A_{p}^{1/3}} 
\]

Let $\overline{m}$ be the average multiplicity produced in N-N collision and 
$\overline{m}_{NA}$ is the corresponding figure for N-A collision at the
same incident energy. So that the average multiplicity produced in A-A
collisions may be calculated in base of N-N as,

\begin{equation}
\overline{m}_{AA}=\nu _{NN}\overline{m}
\end{equation}

and in base of N-A as,

\begin{equation}
\overline{m}_{AA}=\nu _{NA}\overline{m}_{NA}
\end{equation}

Table (1) shows the values of $\overline{m}_{AA}$ as calculated by Eqs.(7 \&
8) compared with the experimental data. The comparison of the last three
columns shows that the predicted values of $\overline{m}\nu _{NN}$ are more
close to the experimental data which implies that the A-A collision behaves
as a base of N-N not N-A collision.

\begin{center}
{\Large Table (1)}

\begin{tabular}{|c|c|c|c|c|c|}
\hline
Reaction & $\nu _{NN}$ & $\nu _{NA}$ & $\overline{m}\ \nu _{NN}$ & $%
\overline{m}_{NA}$ $\nu _{NA}$ & $\overline{m}_{AA}(\exp )$ \\ \hline
p-S & 1.83 & 0.25 & 5.87 & 1.48 & 5.90 \\ \hline
p-Au & 4.20 & 0.42 & 13.5 & 4.14 & 9.90 \\ \hline
d-Au & 7.8 & 0.83 & 25.2 & 11.3 & 23.0 \\ \hline
O-Au & 45.3 & 6.70 & 145.0 & 90.0 & 137.0 \\ \hline
S-S & 25.4 & 8.0 & 81.3 & 47.2 & 90.0 \\ \hline
S-Ag & 54.8 & 27.0 & 175.0 & 160.0 & 162.0 \\ \hline
S-Au & 77.9 & 49.2 & 250.0 & 290.0 & 225.0 \\ \hline
\end{tabular}
\end{center}

\section{The Multi-Peripheral Model}

\subsection{Hadron-Nucleon Collision}

The multi-peripheral technique was found to be a good tool in dealing with
the problems of multi particle production [13,14] in hadron proton (h-p)
interactions. In this technique, the many body-system is expanded into
subsystems, each concerns a two body collision. It is assumed that each
hadron in the final state is produced at a specific peripheral surface that
is characterized by a peripheral parameter. The phase space integral $%
I_{n}(s)$ of the produced hadrons is a measure of the probability of
producing n particle in the final state at center of mass energy $\sqrt{s}$.
It depends mainly on the volume in phase space and the transition matrix
element $T$, and defined as,

\begin{equation}
I_{n}(s)=\int \cdot \cdot \cdot \int \prod_{i}^{n}\frac{d^{3}p_{i}}{2E_{i}}%
\delta ^{4}(\sqrt{s}-\sum_{j}p_{j})|T|^{2}
\end{equation}

Eq. (9) may be simplified as if expressed as a sequence of two particle
decay Fig.(5). Accordingly, the integration in Eq. (9) becomes,

\begin{equation}
I_{n}(s)=\int_{\mu
_{n-1}^{2}}^{(M_{n}-m_{n})^{2}}dM_{n-1}^{2}I_{2}(k_{n}^{2},k_{n-1}^{2},p_{n}^{2})I_{n-1}(M_{n-1}^{2})|T|^{2}
\end{equation}

\[
\ \qquad \;\ \qquad \qquad =\int_{\mu
_{n-1}^{2}}^{(M_{n}-m_{n})^{2}}dM_{n-1}^{2}\int d\Omega _{n-1}\frac{\lambda
^{1/2}(M_{n}^{2},M_{n-1}^{2},m_{n}^{2}}{8M_{n}^{2}}\
I_{n-1}(M_{n-1}^{2})|T|^{2} 
\]

Where $M_{i}$ is the invariant mass of the subsystem composed of $i-$%
particles. The integration over all possible values of $M_{n-1}$ concerns
the first vertex in the chain of Fig.(5). To proceed further, we iterate
Eq.(10) for the $M_{n-2},M_{n-3},M_{2}$ to obtain the entire chain, so that,

\[
I_{n}(s)=\int_{\mu _{n-1}}^{(M_{n}-m_{n})}dM_{n-1}d\Omega _{n-1}\frac{1}{2}%
P_{n-1}|T(P_{n-1})|^{2}\cdot \cdot \cdot \int_{\mu
_{2}}^{(M_{3}-m_{3})}dM_{2}d\Omega _{2}\frac{1}{2}P_{3}|T(P_{2})|^{2} 
\]

\begin{equation}
\cdot \int d\Omega _{1}\frac{1}{2}P_{2}|T(P_{1})|^{2}
\end{equation}

Where $P_{i}=\lambda ^{1/2}(M_{i}^{2},M_{i-1}^{2},m_{i}^{2})$ is the
three-vector momentum of the ith particle and $T(p_{i})$ is the transition
matrix element at the i$^{th}$ vertex. For the case of strong interactions $%
T(p_{i})$ has a parametric form which cut the magnitude of the particles
momenta according to their physical region [16,19],

\begin{equation}
T(p_{i})=\exp (-\alpha _{i} p_{i})
\end{equation}

And $\alpha _{i}$ is a peripheral parameter, which fits the experimental
data. The multiple integration in Eq. (11) may be solved by the Monte Carlo
technique [20-22]. At extremely high energy, Eq. (11) has an asymptotic
limit in the form;

\begin{equation}
I_{n}(s)=\frac{(\pi /2)^{n-1}}{(n-1)!(n-2)!}s^{n-2}|T|^{2(n-1)}
\end{equation}

\subsection{Hadron-Nucleus Collisions}

On extending the model to the hadron-nucleus or nucleus-nucleus collisions,
we follow the NN-base super position as expected from the features of the
experimental data that are illustrated in section {\bf 2}. We should
consider the possible interactions with the nucleons forming the target
nucleus $A_{t}$. The incident hadron makes successive collisions inside the
target. The energy of the incident hadron (leading particle) slows down
after each collision, producing number of created hadrons each time that
depends on the available energy. The phase space integral $I_{n}^{NA}$ in
this case has the form,

\begin{equation}
I_{n}^{NA}(s)=\sum_{\nu }^{A_{t}}I_{n_{\nu }}(s_{\nu })P(\nu ,A_{t})\delta
(n-\sum_{i}^{A_{t}}n_{i})
\end{equation}

Where $P(\nu ,A_{t})$ is the probability that $\nu $ nucleons out of $A_{t}$
will interact with the leading particle and $I_{n_{\nu }}(s_{\nu })$ is the
phase space integral of $NN$ collision that produces hadrons at energy $%
\sqrt{s_{\nu }}$. The delta function in Eq.(14) is to conserve the number of
particle in the final state. Treating all nucleons identically, and that $%
X_{NN}$ is the $N-N$ phase shift function [18,23-25], then, according to the
eikonal approximation,

\begin{equation}
P(l,A_{t})=-\left( 
\begin{array}{c}
A_{t} \\ 
l
\end{array}
\right) \sum_{j=0}^{l}(-1)^{j}\left( 
\begin{array}{c}
l \\ 
j
\end{array}
\right) \{1-\exp (2Re i(A_{t}-l+j)X_{NN})\}
\end{equation}

The working out of this approach is to put the multi-dimension integration
of Eq. (11) and the generated kinematical variables into a Monte Carlo
subroutine [16]. This in turn is restored $\nu $ times, where $\nu $ is the
number of collisions inside the target nucleus that is generated by a Monte
Carlo Generator according to the probability distribution Eq.(15). In the
first one, the incident hadron has its own incident energy $E_{0}$ and moves
parallel to the collision axis (z-axis) $\theta _{0}=0$. The output of the
subroutine determines the number of created hadrons n1 as well as the energy 
$E_{1}(<E_{0})$ and the direction q1 of the leading particle. The leading
particle leads the reaction in its second round with $E_{1}$ and $\theta
_{1} $ as input parameters and creates new number of particles $n-2$ and so
on. The number nj is determined according to a multiplicity generator which
depends on the square of the center of mass energy sj in the round number j.

\begin{equation}
s_{j}=2m_{N}^{2}+2m_{N}E_{j}
\end{equation}

Fig. (6) demonstrates the particle rapidity distribution produced in the
first three- collisions as predicted by the model for the p- $^{32}$S at 200
GeV. The yield (as measured by the area under the curve) decreases with the
order of the collision because of the appreciable drop in the energy after
the successive collisions. On the other hand, the family of curves
concerning the different number of collisions acquires gradually decreased
rapidity range. The overall distribution is compared with the experimental
data SLAC-NA-035 in Fig. (7). The comparison shows good agreement.

\subsection{Nucleus-Nucleus Collisions}

The extension of the multi peripheral model to the nucleus- nucleus case is
more complicated. The number of available collisions is multi-folded due to
the contribution of the projectile nucleons. By analogy to the N-A
collision, it is possible to define the phase space integral $I_{n}^{AA}$ in 
$A-A$ collisions as,

\begin{equation}
I_{n}^{AA}(s)=\sum_{j}^{A_{p}}%
\sum_{k}^{A_{t}}I_{n_{j,k}}(s_{j,k})P_{AA}(j,A_{p},k,A_{t})\delta
(n-\sum_{j,k}^{A_{t}}n_{j,k})
\end{equation}

Where $I_{n_{j,k}}(s_{j,k})$ is the phase space integral due to the knocked
on nucleon number $j$ from the projectile and that, number $k$ from the
target. The probability that the $A-A$ collision encounters $\nu _{p}$
collisions from the projectile and $\nu _{t}$ collisions from the target is
treated as independent events. So that,

\begin{equation}
P_{AA}(j,A_{p},k,A_{t})=P(\nu _{p},A_{p})\cdot P(\nu _{t},A_{t})
\end{equation}

Another modification is carried out on this calculation that is to consider
the screening effect of the projectile nucleons upon the interaction. This
effect is summarized as follows. The first projectile nucleon faces the
target nucleus as a whole, i.e. it can see the complete $A_{t}$ nucleons and
interacts with only $\nu _{t1}$ of them. The next projectile nucleon will
see that target as partially screened by the first. The target size in this
case is $A_{t}-\nu _{t1}$ and it interacts with only $\nu _{t2}$ nucleons
according to a probability function, $P(\nu _{t2},A_{t}-\nu _{t1})$ , by
simple iteration, the i$^{th}$ projectile nucleon will see a target as $%
A_{t}-\sum_{k}^{i-1}\nu _{tk}$and so on. The prediction of the model is
applied to the $^{32}$S-$^{32}$S collisions at 200 A GeV incident energy.
The rapidity distribution of the produced particles in the first 3-
projectile nock on nucleons is demonstrated in Fig. (8). While the overall
rapidity distribution is compared with the data of the experiment
SLAC-NA-035 in Fig. (9). The fair agreement obtained for $N-A,$ Fig. (7) and 
$A-A$ collisions Fig. (9) supports the success of the multi-peripheral
model. A small defect in the model for the $NN$ case will be multi-folded
and would be seen clearly in $N-A$ and $A-A$ collisions.

\section{Summary and Conclusive Remarks}

\qquad 1- A scaling rapidity function is obtained for particles produced in
h-N, $A-A$ and $A-A$ collisions, which assumes that the reaction is built on 
$N-N$ base.

2- The multi-peripheral model is extended to the nucleon-nucleus and the
nucleus-nucleus interactions on bases of nucleon-nucleon collisions.

3- The phase space integral of the nucleon-nucleon collision is folded
several times according to the number of encountered nucleons from the
target.

4- The probability that $\nu $ nucleons from the target are encountered by a
projectile nucleon is calculated in terms of the nucleon-nucleon phase shift
according to the eikonal approximation.

5- The number of created particles in each collision is summed over to get
the production in the nucleon-nucleus case, where the conservation of number
of particles in the final state is taken into consideration.

6- In nucleus-nucleus collisions, we followed the statistics of independent
events. The screening effect among the interacting projectile nucleons is
also considered.

\newpage
{\Large Figure Captions}

Fig. (1) The rapidity distribution of particles produced in nucleus 
collisions at the same incident energy.\\ 

Fig. (2) The multiplicity multiplication factor in nucleus-nucleus 
collisions in terms of the yield in pp collision at the same incident energy. 
The p-S collision is represented by the black- plus line, S-S (crossed), S-
Ag (triangle) and S-Au (circles). \\

Fig. (3) The multiplicity multiplication factor in nucleus-nucleus 
collisions in terms of the yield in p-S collision at the same incident 
energy. The black- plus line represents the S-S collision, S-Ag (crossed) 
and S-Au (triangles). \\

Fig. (4) The scaled rapidity distribution produced from the nucleus-
nucleus collisions at the same incident energy. \\

Fig. (5) A tree diagram for the multi-peripheral process, producing n-
particles in the final state.  \\

Fig. (6) The particle rapidity distribution produced in p-S at multiple 
order collisions as predicted by the MPM. \\

Fig. (7) The particle rapidity distribution produced in p-S collision at 200 
GeV incident proton energy as predicted by the MPM and compared with 
the experimental data SLAC-NA-035.\\ 

Fig. (8) The particle rapidity distribution produced in S-S at multiple 
order collisions as predicted by the MPM.\\

Fig. (9) The particle rapidity distribution produced in S-S collision at 200 
GeV incident proton energy as predicted by the MPM and compared with 
the experimental data SLAC-NA-035. \\

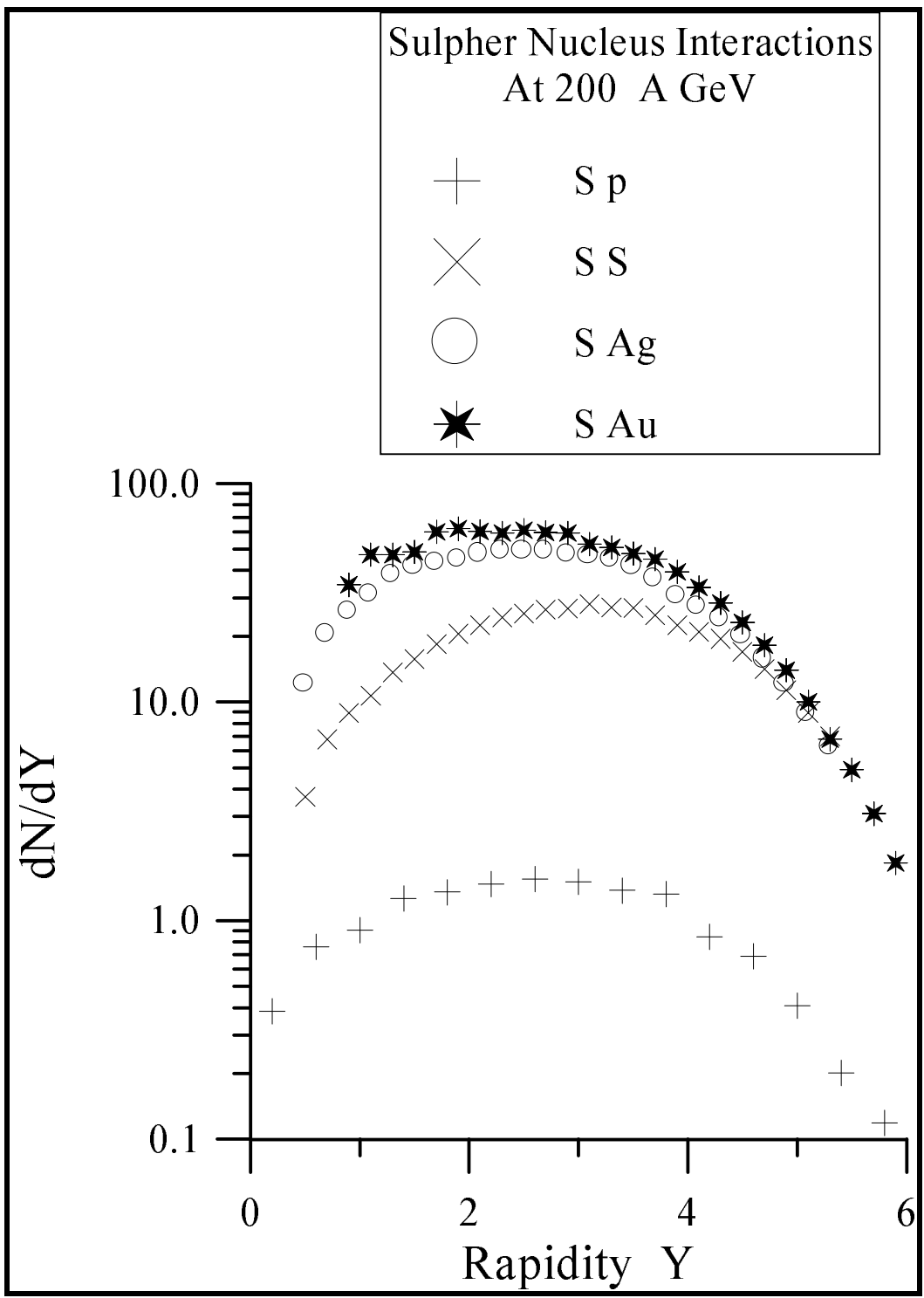
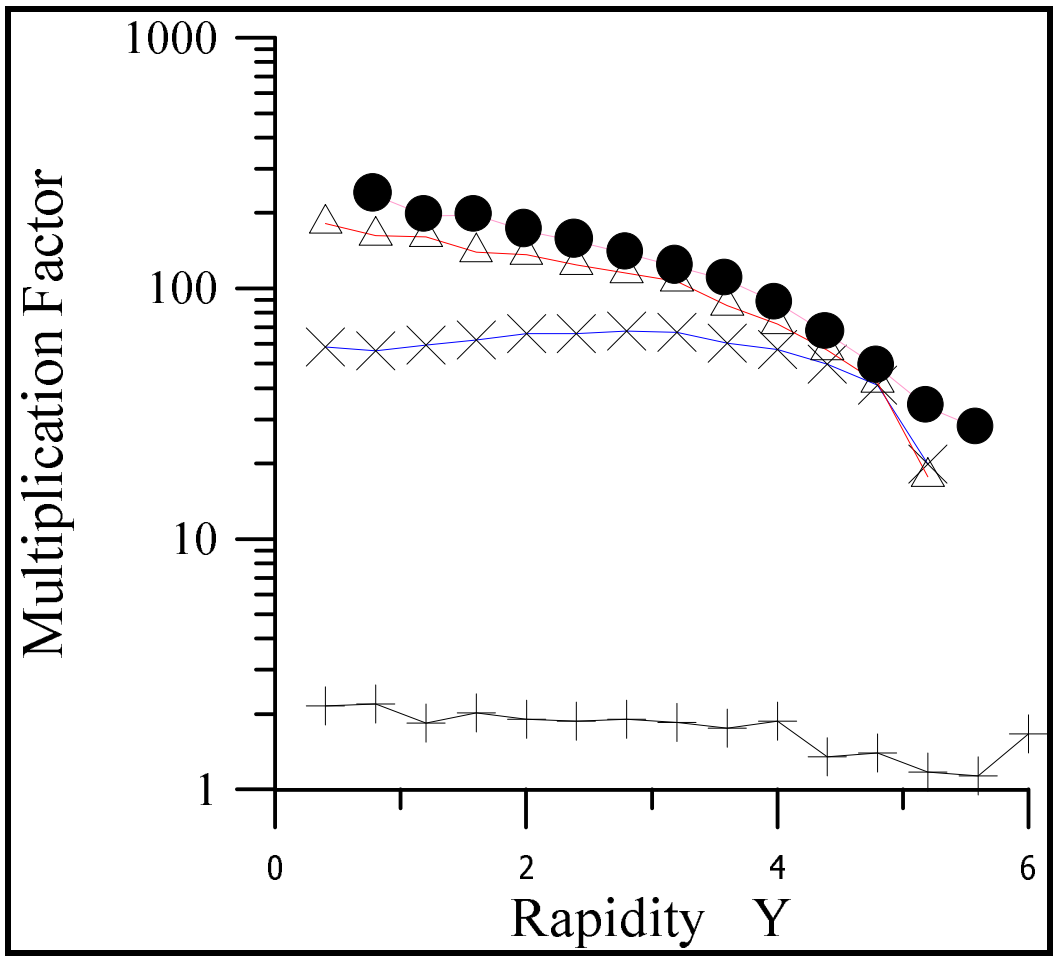
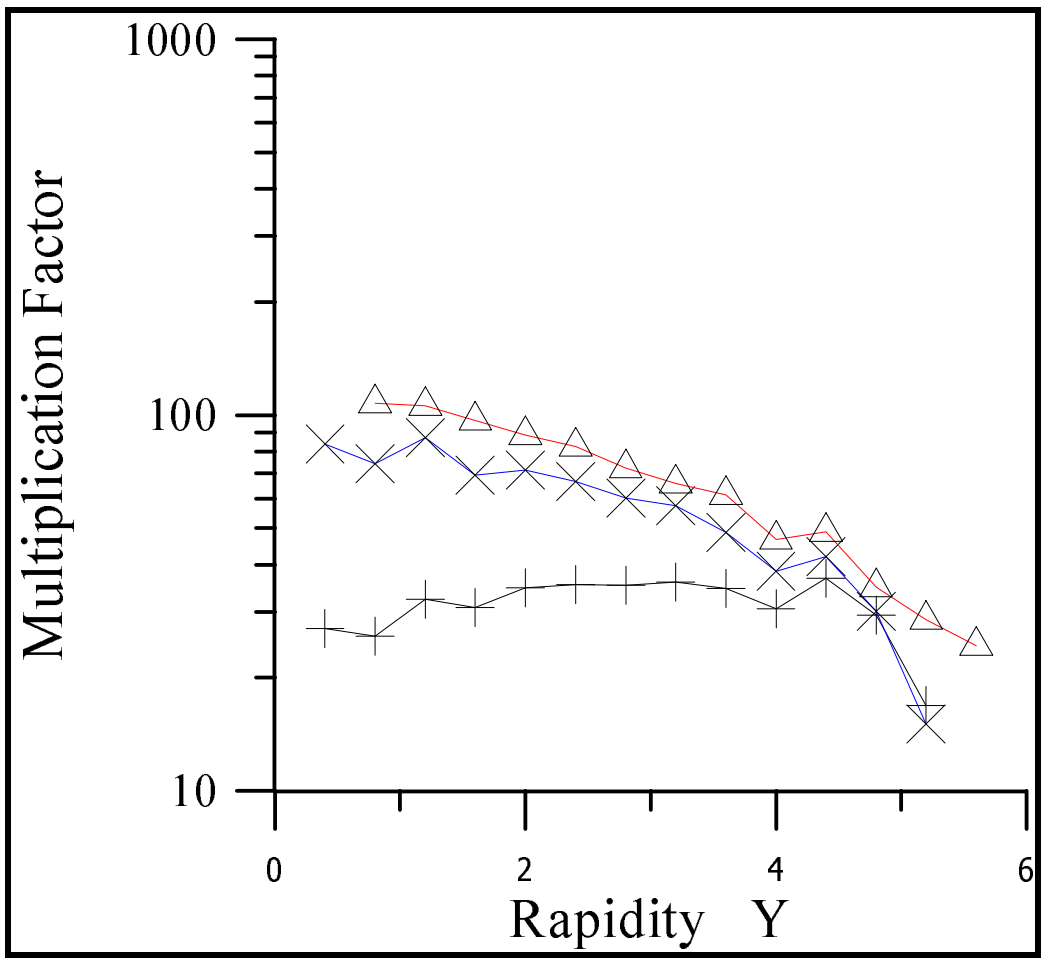
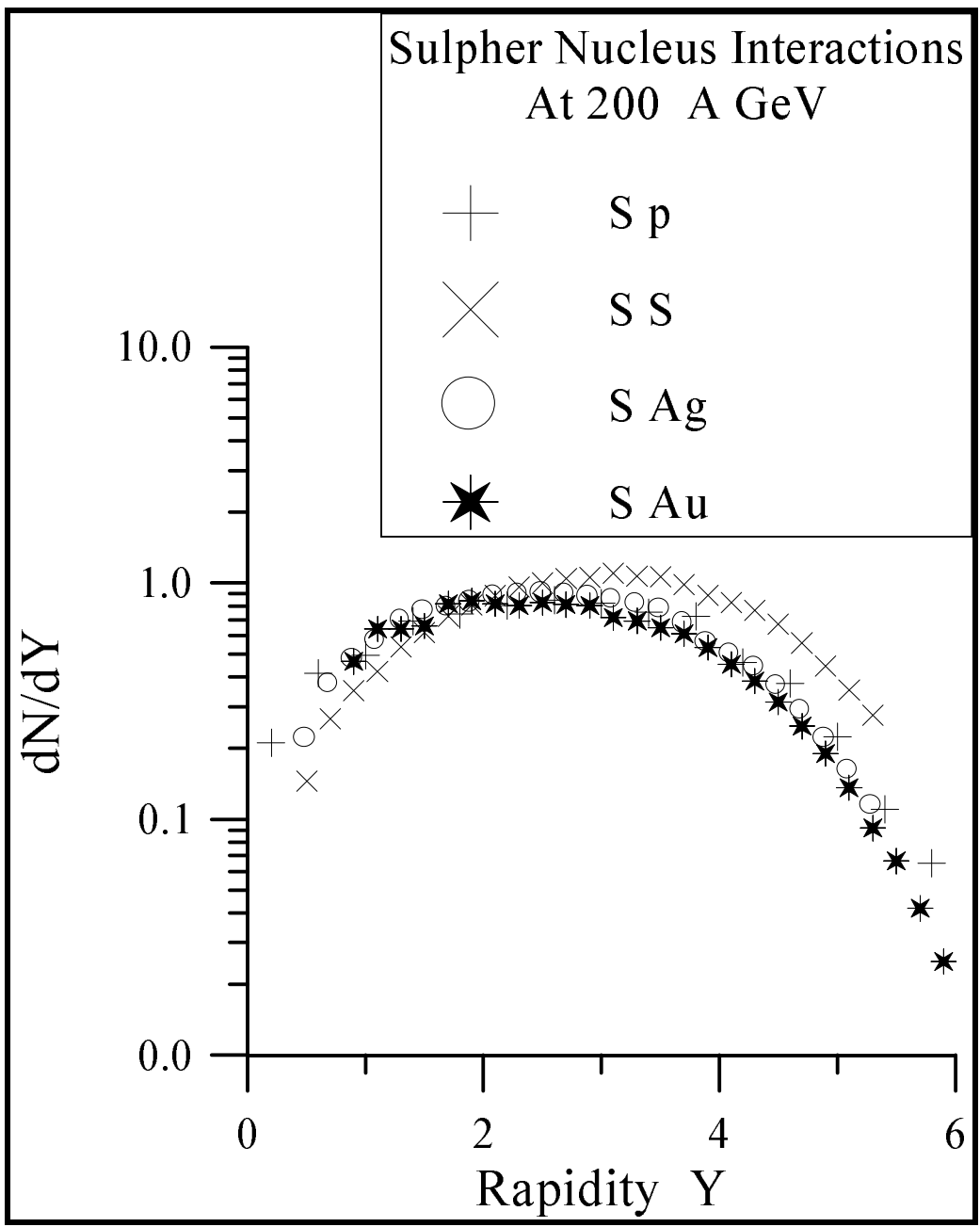
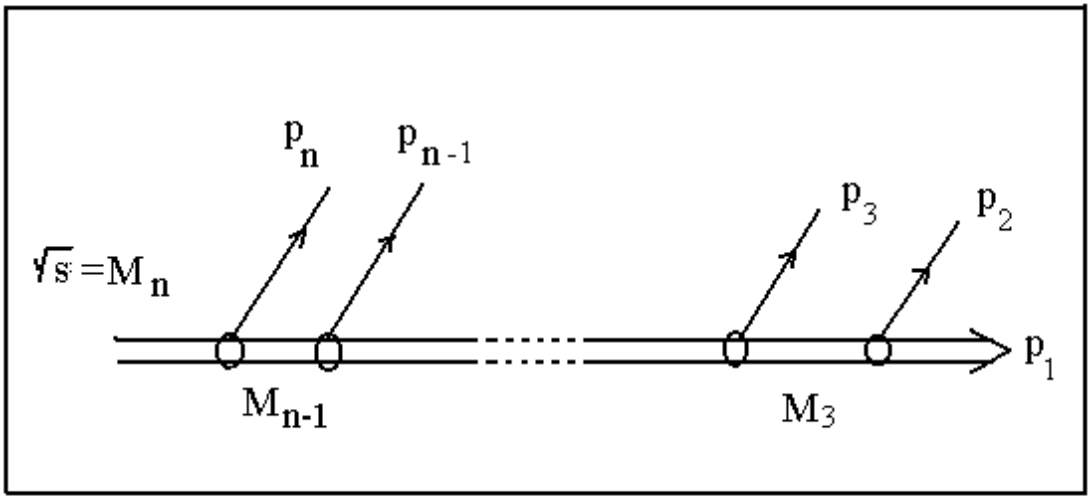
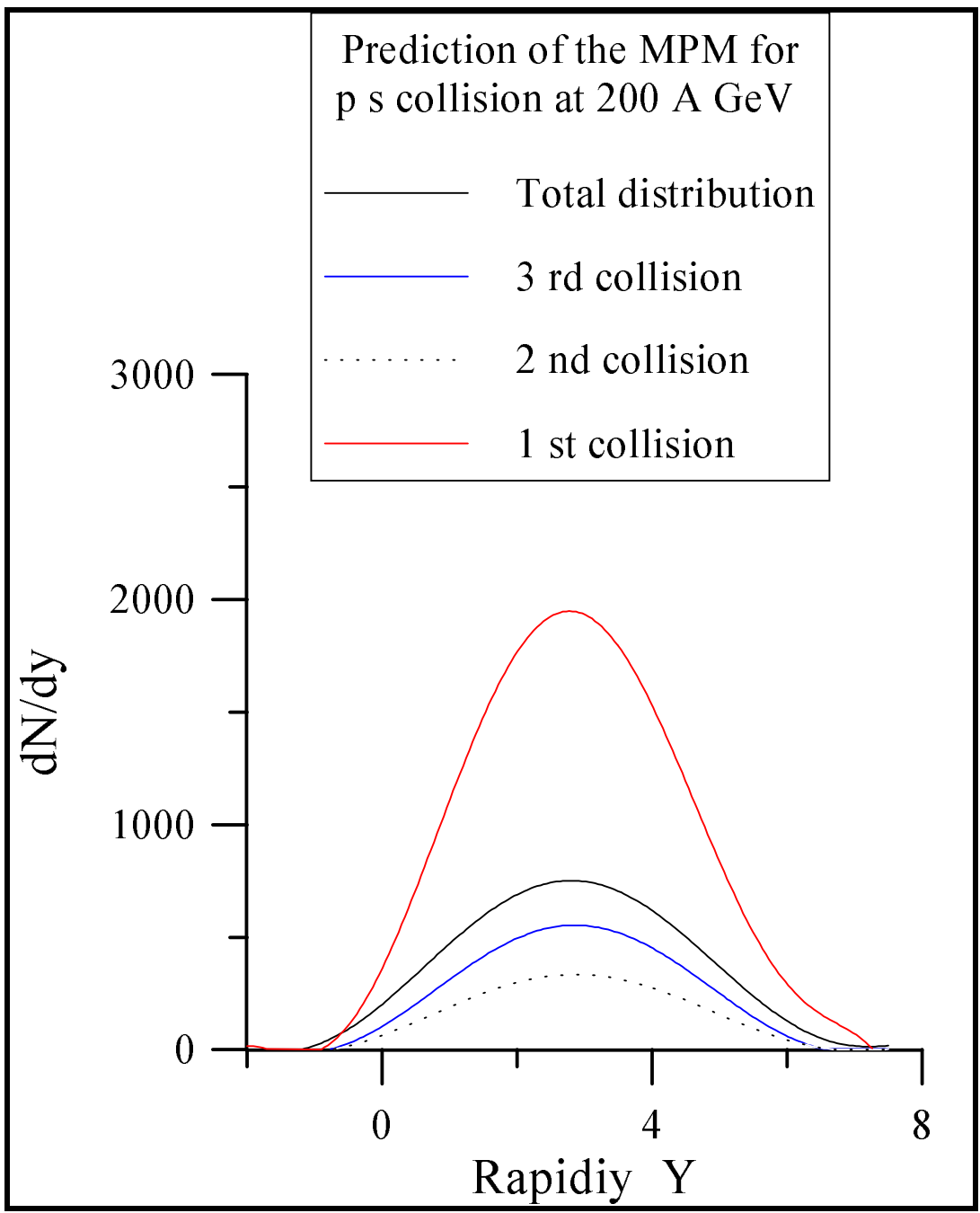
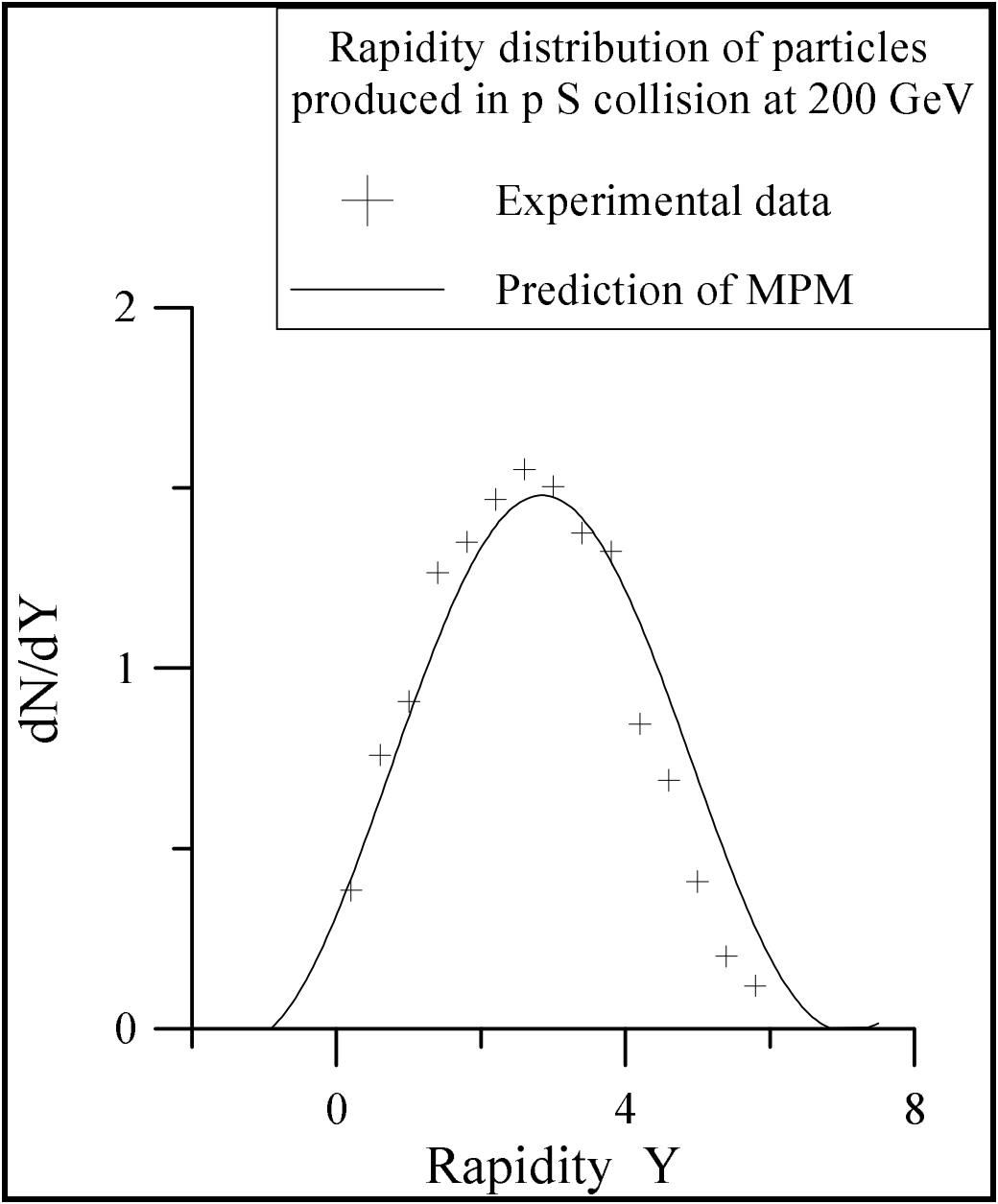
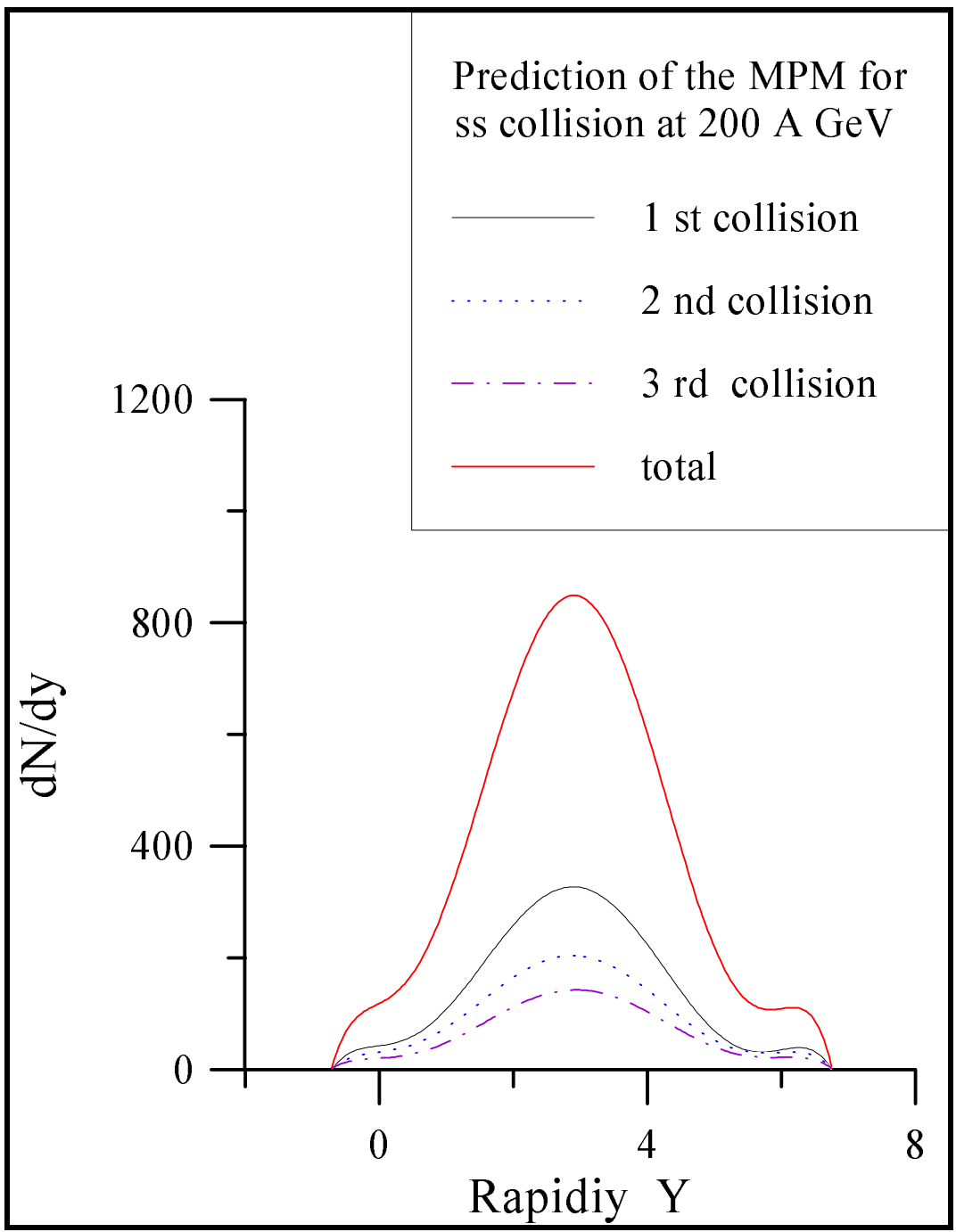
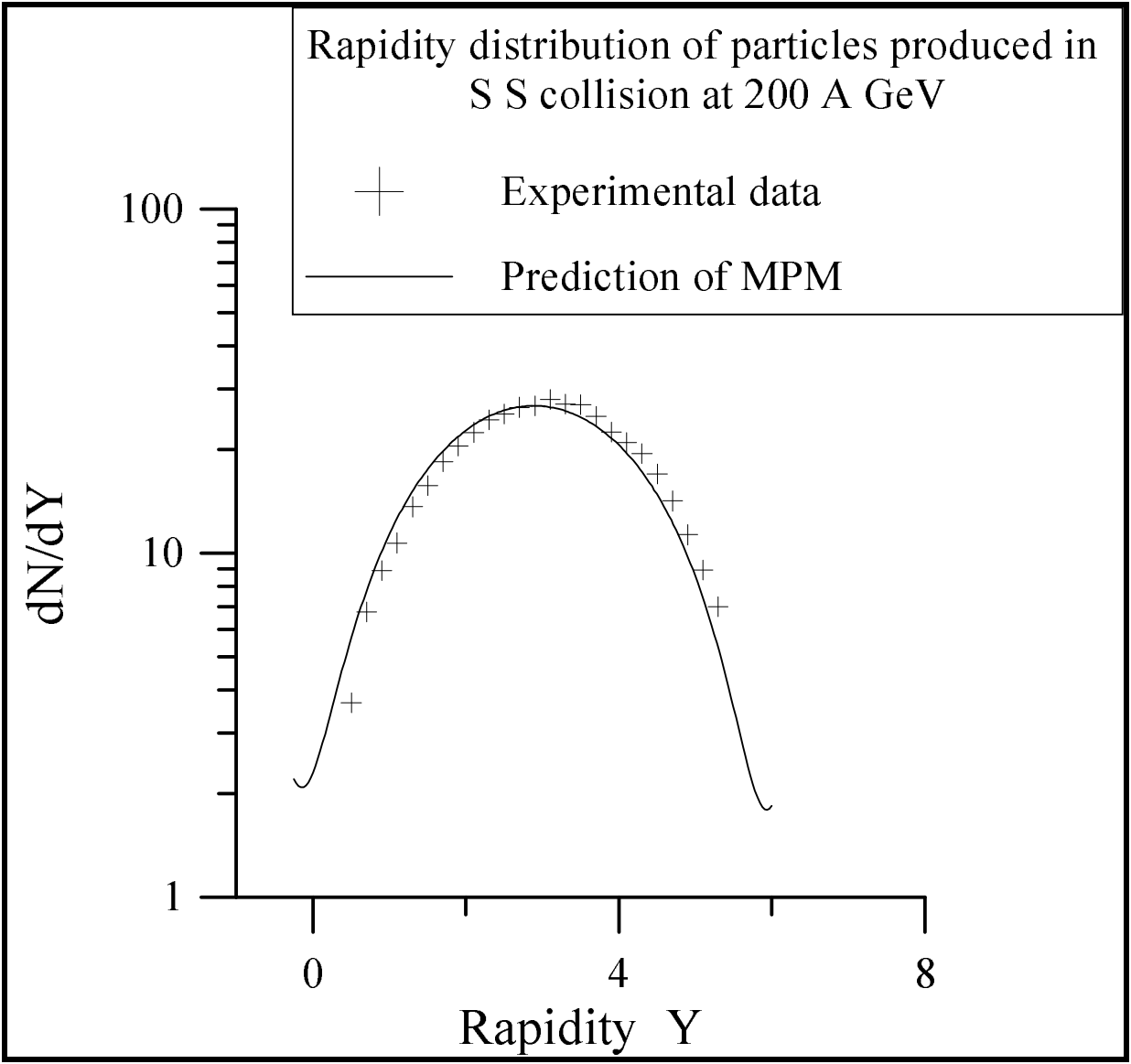
\end{document}